\documentclass{svmult}

\usepackage{amsmath}
\usepackage{amsfonts}
\usepackage{amssymb}
\usepackage{makeidx}         
\usepackage{graphicx}        
\usepackage{multicol}        

\makeindex             

\begin{document}

\title*{Identifying Excessively Rounded\\ or Truncated Data}
\author{Kevin H. Knuth\inst{1}\and
J. Patrick Castle\inst{2}\and Kevin R. Wheeler\inst{3}}
\institute{Department of Physics, University at Albany, Albany NY
12222\\
\texttt{kknuth@albany.edu} \and Intelligent Systems
Division, NASA Ames Research Center,\\ Moffett Field CA 94035\\
\texttt{pcastle@email.arc.nasa.gov} \and Monterey Bay Aquarium
Research Institute, Moss Landing CA 95039\\
\texttt{kevinwheeler@ieee.org}}
%
%
\maketitle

\textbf{Abstract.} All data are digitized, and hence are
essentially integers rather than true real numbers.  Ordinarily
this causes no difficulties since the truncation or rounding
usually occurs below the noise level.  However, in some instances,
when the instruments or data delivery and storage systems are
designed with less than optimal regard for the data or the
subsequent data analysis, the effects of digitization may be
comparable to important features contained within the data.  In
these cases, information has been irrevocably lost in the
truncation process. While there exist techniques for dealing with
truncated data, we propose a straightforward method that will
allow us to detect this problem before the data analysis stage. It
is based on an optimal histogram binning algorithm that can
identify when the statistical structure of the digitization is on
the order of the statistical structure of the data set itself.

\section{Data}
\label{sec:1} All data are digitized, whether it is a number
written in a researcher's lab notebook or a measurement recorded
and stored in a robotic explorer's memory system.  This fact, by
itself, is not surprising or unexpected, since it is impossible to
physically express all real numbers with complete precision.
Though what is perhaps surprising, is that this fact can have
unforeseen consequences, especially as the truncation level
approaches the noise level \cite{Bayman&Broadhurst}.  In these
cases, the digitization effect can occlude or eradicate important
structure in the data themselves.

With the impressive advances in Bayesian inferential technology,
we have found that relevant information can be retrieved from data
well below what we have traditionally believed to be the noise
floor.  Since our computational technology has advanced beyond the
point envisioned by many instrument designers, it is possible that
current, and planned, instruments are not designed to return data
with the precision necessary for the most modern of our
computational techniques.

In this paper, we propose a straightforward method that allows us
to identify situations where the data have been excessively
truncated or rounded.  This method relies on constructing the
simplest of models of the data---a density function, which is
simplified further by modelling the density function as a
piecewise-constant function. Relying on Bayesian probability
theory to identify the optimal number of bins comprising the
density model, we can identify situations where the information
contained in the data is compromised by the digitization effect.

\section{Density Models}
\label{sec:2} One of the simplest ways of describing data is to
describe the range of values it can take along with the
probabilities with which it takes those values.  Such models are
called density models.  To this day, the most commonly used form
of density model is a histogram, where the range of values is
divided into a number of bins $M$ and the bin heights are
determined by the number of data points that fall within the bin.
The bin probability is easily computed from the number of data
points within the bin divided by the total number of data points.

Other commonly used density models are kernel density estimators,
which introduces a narrowly peaked probability density function at
each datum point and sums each of these functions to obtain the
entire density function.  If one possesses sufficient prior
knowledge to know the functional form of the density function,
such as that it is a Gaussian distribution, one only needs to
estimate the parameters of that distribution from the data.  In
the case of the Gaussian distribution, we need to estimate $\mu$
and $\sigma$.

\subsection{The Piecewise-Constant Model}
\label{sec:2.1} For the sake of simplicity, we choose to model the
density function with a piecewise-constant model.  A histogram can
be viewed as a piecewise-constant model, although it is not
properly normalized. We shall show greater care in our treatment.

We begin by dividing the range of values of the variable into a
set of $M$ discrete bins and assigning a probability to each bin.
We denote the probability that a datum point is found to be in the
$k^{th}$ bin by $\pi_k$.  Since we require a density function, we
require that the ``height'' of the bin $h_k$ be dictated by the
\emph{probability density} of the bin, which is the probability of
the bin divided by its width $v_k$.  This gives
\begin{equation}
h_k = \frac{\pi_k}{v_k}.
\end{equation}
Integrating this constant probability density $h_k$ over the width
of the bin $v_k$ leads to a total probability $\pi_k = h_k v_k$
for the bin. This leads to the following piecewise-constant model
$h(x)$ of the unknown probability density function for the
variable $x$
\begin{equation}
h(x) = \sum_{k = 1}^{M}{h_k~\Pi(x_{k-1}, x, x_k)},
\end{equation}
where $h_k$ is the probability density of the $k^{th}$ bin with
edges defined by $x_{k-1}$ and $x_k$, and $\Pi(x_{k-1}, x, x_k)$
is the boxcar function where
\begin{equation}
\Pi(x_a, x, x_b) =
    \left\{ \begin{array}{rl}
       0 & \mbox{if}~~x < x_a\\
       1 & \mbox{if}~~x_a \leq x < x_b\\
       0 & \mbox{if}~~x_b \leq x \end{array} \right.
\end{equation}
If equal bin widths are used, the density model can be re-written
in terms of the bin probabilities $\pi_k$ as
\begin{equation}
h(x) = \frac{M}{V} \sum_{k = 1}^{M}{\pi_k~\Pi(x_{k-1}, x, x_k)}.
\end{equation}
where $V$ is the width of the entire region covered by the density
model.

\subsection{Bayesian Probability Theory}
By applying Bayesian probability theory
\cite{Sivia:1996,Gelman+etal:1995} we can use the data to
determine the optimal or expected values of the model parameters,
which are the number of bins $M$ and the bin probabilities
$\underline{\pi} = \{\pi_1, \pi_2, \ldots, \pi_{M-1}\}$.  Bayes'
Theorem states that
\begin{equation}
p(model | data, I) \propto p(model | I) \cdot p(data | model, I),
\end{equation}
where the symbol $I$ is used to represent any prior information
that we may have or any assumptions that we have made, such as the
assumption that the bins are of equal width.  The probability on
the left $p(model|data,I)$ is called the posterior probability,
which describes the probability of a set of particular values of
the model parameters given both the data and our prior
information. The first probability on the right $p(model|I)$ is
called the prior probability since it describes the probability of
the model parameter values before we have collected any data.  The
second probability on the right $p(data|model,I)$ is called the
likelihood since it describes the likelihood that the observed
data could have been generated by the model.  The inverse of the
implicit proportionality constant is called the evidence.  In this
paper, it will not be necessary to compute this quantity as long
as we are content to work with posterior probabilities which have
not been normalized so that their sum is equal to one.

If we write the observed data as $\underline{d} = \{d_1, d_2,
\ldots, d_N\}$, Bayes' Theorem becomes
\begin{equation}
p(\underline{\pi}, M | \underline{d} ,I) \propto
p(\underline{\pi}, M | I) \cdot p(\underline{d} | \underline{\pi},
M, I),
\end{equation}
where the joint prior can be further decomposed using the product
rule
\begin{equation}
p(\underline{\pi}, M | \underline{d} ,I) \propto p(\underline{\pi}
| M, I) \cdot p(M | I) \cdot p(\underline{d} | \underline{\pi}, M,
I).
\end{equation}
Next we must assign functions for the likelihood and the two prior
probabilities.

First, we assign the likelihood to be the multinomial likelihood
\begin{equation} \label{eq:likelihood}
p(\underline{d} | \underline{\pi}, M, I) =
\biggl(\frac{M}{V}\biggr)^N \pi_1^{n_1} \pi_2^{n_2} \ldots
\pi_{M-1}^{n_{M-1}} \pi_{M}^{n_M},
\end{equation}
where $n_i$ is the number of data points in the $i^{th}$ bin.

Second, we assign a uniform prior probability for the number of
bins $M$ defined over a range $0 < M \leq C$
\begin{equation} \label{eq:prior-for-M}
p(M | I) =
   \left\{ \begin{array}{rl}
       C^{-1} & \mbox{if}~~1 \leq M \leq C\\
       0 & \mbox{otherwise} \end{array} \right.
\end{equation}
where $C$ is the maximum number of bins to be considered. This
could reasonably be set to the range of the data divided by
smallest non-zero distance between any two data points.

Last, we assign a non-informative prior for the bin parameters
$\pi_1, \pi_2, \ldots, \pi_{M-1}$
\begin{equation} \label{eq:prior-for-pi's}
p(\underline{\pi} | M, I) =
\frac{\Gamma\bigl(\frac{M}{2}\bigr)}{\Gamma\bigl(\frac{1}{2}\bigr)^M}
\biggl[\pi_1 \pi_2 \cdots \pi_{M-1}
\biggl(1-\sum_{i=1}^{M-1}{\pi_i}\biggr)\biggr]^{-1/2}.
\end{equation}
This is known as the Jeffreys's prior for the multinomial
likelihood (\ref{eq:likelihood})
\cite{Jeffreys:1961,Box&Tiao:1992,Berger&Bernardo:1992}, which has
the advantage in that it is also the conjugate prior to the
multinomial likelihood.

The posterior probability of the model parameters
\cite{Knuth+Gotera+Curry+etal:2005,Knuth:optBINS} is then written
as
\begin{align} \label{eq:joint-posterior}
p(\underline{\pi}, M | \underline{d}, I)& \propto
\biggl(\frac{M}{V}\biggr)^N
\frac{\Gamma\bigl(\frac{M}{2}\bigr)}{\Gamma\bigl(\frac{1}{2}\bigr)^M}~~~\times\\
& \pi_1^{n_1-\frac{1}{2}} \pi_2^{n_2-\frac{1}{2}} \ldots
\pi_{M-1}^{n_{M-1}-\frac{1}{2}}
\biggl(1-\sum_{i=1}^{M-1}{\pi_i}\biggr)^{n_M-\frac{1}{2}},\nonumber
\end{align}
where $1 \leq M \leq C$ and $C^{-1}$ has been absorbed into the
implicit proportionality constant.

\subsection{Estimating the Density Parameters}
\label{sec:2.2} We can obtain the marginal posterior probability
of the number of bins given the data by integrating over all
possible bin heights
\cite{Knuth+Gotera+Curry+etal:2005,Knuth:optBINS}. These $M-1$
integrations results in
\begin{equation} \label{eq:posterior-for-M}
p(M | \underline{d}, I) \propto \biggl(\frac{M}{V}\biggr)^N
\frac{\Gamma\bigl(\frac{M}{2}\bigr)}{\Gamma\bigl(\frac{1}{2}\bigr)^M}~
\frac{\prod_{k=1}^{M}{\Gamma(n_k+\frac{1}{2})}}{\Gamma(N+\frac{M}{2})},
\end{equation}
where the $\Gamma(\cdot)$ is the Gamma function
\cite[p.~255]{Abramowitz&Stegun}. To find the optimal number of
bins, we evaluate this posterior probability for all the values of
the number of bins within a reasonable range and select the result
with the greatest probability. In practice, it is often much
easier computationally to work with the logarithm of the
probability, (\ref{eq:posterior-for-M}) above.  It is important to
note that the equation above is a proportionality, which means
that there is a missing proportionality constant.  Thus the
resulting posterior is not normalized to have a value between zero
and one.

\begin{figure}
\centering
\includegraphics[width=0.85\textwidth]{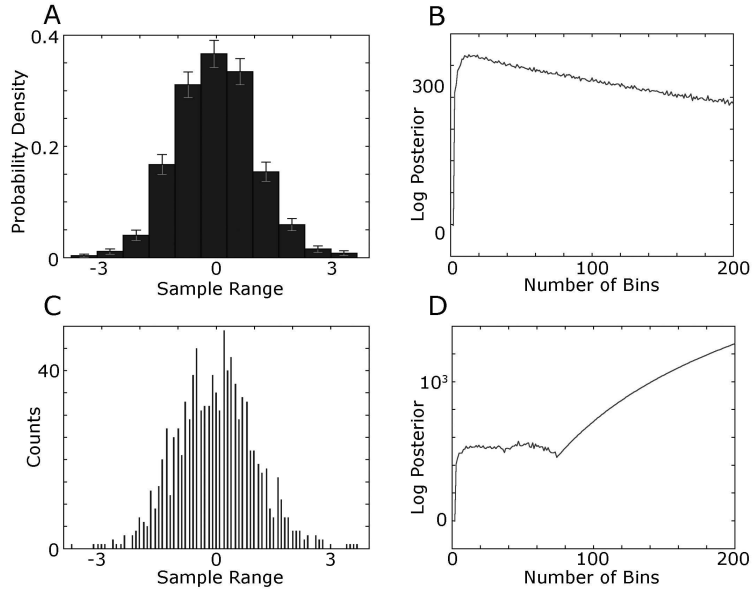}
%
%
\caption{In this example we take 1000 data points sampled from a
Gaussian distribution. A) Here we show the optimal
piecewise-constant model with $M = 11$ bins. The bin heights
represent the probability density within the region bounded by the
bin edges, and the error bars represent one standard of deviation
of uncertainty about the estimated value for the probability
density. B) This figure shows the logarithm of the un-normalized
marginal posterior probability density for the number of bins.
Generally the log posterior rises sharply rounding off to a peak
and then falls of gently as $M$ increases.  In this case, the peak
occurs at $M = 11$ indicating the optimal number of bins. C) We
took the same 1000 data points, but rounded their values to the
nearest 1/10th. The optimal solution looks like a picket fence
highlighting the discrete nature of the data rather than the
Gaussian nature.  D) The un-normalized log posterior rises sharply
as before, but does not indicate an optimal peak. As $M$ increases
and the discrete data can be separated no further, the log
posterior changes behavior and increases asymptotically to a value
greater than zero.  This is a clear indication of the discrete
nature of the data due to excessive rounding.}
\label{fig:1}       
\end{figure}

Once the number of bins have been selected, we can use the joint
posterior probability (\ref{eq:joint-posterior}) to compute the
mean bin probabilities and the standard deviations of the bin
probabilities from the data
\cite{Knuth+Gotera+Curry+etal:2005,Knuth:optBINS}.  The mean bin
probability is
\begin{equation} \label{eq:mean-prob}
\mu_k = \langle h_k \rangle = \frac{\langle \pi_k \rangle}{v_k} =
\biggl(\frac{M}{V}\biggr)
\biggl(\frac{n_k+\frac{1}{2}}{N+\frac{M}{2}}\biggr),
\end{equation}
and the associated variance of the height of the $k^{th}$ bin is
\begin{equation} \label{eq:var-prob}
\sigma_k^2 = \biggl(\frac{M}{V}\biggr)^2
\biggl(\frac{(n_k+\frac{1}{2})(N-n_k+\frac{M-1}{2})}
{(N+\frac{M}{2}+1)(N+\frac{M}{2})^2}\biggr),
\end{equation}
where the standard deviation is the square root of the variance.
This result again differs from the traditional histogram since
bins with no counts still have a non-zero probability. This is in
some sense comforting, since no lack of evidence can ever prove
conclusively that an event occurring in a given bin is
impossible---just less probable.

These computational methods allow us to estimate probability
densities from data, and quantify the uncertainty in our
knowledge. An example of a probability density model is shown in
Figure \ref{fig:1}A.

Looking at the logarithm of the marginal posterior probability for
the number of bins $M$ (Figure \ref{fig:1}B), we see that it
typically rises rapidly as the likelihood term increases with
increasing numbers of bins. However, as the number of bins becomes
large, the prior probability dominates causing the posterior
probability to decrease.  It is this balance between the
data-driven likelihood and the prior probability that sets up a
region where there is an optimal number of bins.

This optimal binning technique ensures that our density model
includes all the relevant information provided by the data while
ignoring irrelevant details due to sampling variations.  The
result is the most honest summary of our knowledge about the
density function from the given data. We will now look at the
asymptotic behavior of the marginal posterior probability
(\ref{eq:posterior-for-M}) and see how it changes as digitization
in the data becomes relevant information.

\section{Asymptotic Behavior}
\label{sec:3} 
\begin{figure}[t]
\centering
\includegraphics[width=0.5\textwidth]{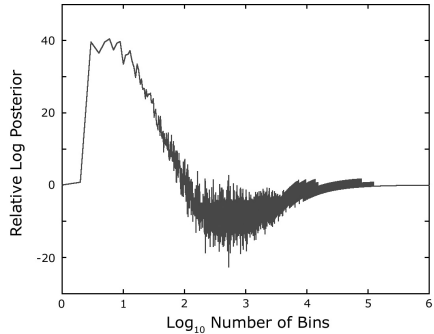}
\caption{In this example we take 200 data points sampled from a
Gaussian distribution and demonstrate the asymptotic behavior of
the log posterior. Note that the x-axis displays the log base 10
of the number of bins.  Note that the function asymptotes to zero
for extremely large numbers of bins.}
\label{fig:2}       
\end{figure}

In the event that we have a number of bins much
greater than the number of data, $M >> N$, where each datum point
is in a separate bin, the marginal posterior probability for $M$
(\ref{eq:posterior-for-M}) becomes
\begin{equation}\label{eq:M>N}
p(M | \underline{d}, I) \propto \biggl(\frac{M}{2}\biggr)^N
\frac{\Gamma\bigl(\frac{M}{2}\bigr)}{\Gamma\bigl(N+\frac{M}{2}\bigr)},
\end{equation}
which can be rewritten as
\begin{equation}
p(M | \underline{d}, I) \propto \biggl(\frac{M}{2}\biggr)^N
\biggl[\biggl(N-1+\frac{M}{2}\biggr)\biggl(N-2+\frac{M}{2}\biggr)\cdots\biggl(\frac{M}{2}\biggr)\biggr]^{-1}.
\end{equation}
Since there are $N$ terms involving $M$ in the product on the
right, the posterior probability can be seen to approach one as $M
\rightarrow \infty$. As expected, Figure \ref{fig:2} shows that
the log posterior approaches zero in that limit.

\subsection{Identifying Excessively Rounded or Truncated Data}
\label{sec:3.1} In the event that the data are digitized it will
be impossible (with sufficient data) for every datum point to be
in its own bin as the number of bins increases.  Specifically, we
can expect that once the bin width has become smaller than the
precision of the data, increasing the number of bins $M$ will not
change the number of populated bins $P$ nor their populations
$n_p$, although it will change \emph{which} bins are populated. If
the precision of the data is $\Delta x$, we define
\begin{equation}
M_{\Delta x} = \frac{V}{\Delta x},
\end{equation}
where $V$ is the range of the data considered.  Now for $M >
M_{\Delta x}$ the number of populated bins $P$ will remain
unchanged since the bin width $w$ for $M > M_{\Delta x}$ will be
smaller than the precision, $w < \Delta x$.

For bin numbers $M > M_{\Delta x}$, there will be $P$ populated
bins with populations $n_1, n_2, \ldots, n_P$. \footnote{We should
be more careful with the indices here, since by varying $M$, the
indices to the particular bins will change. A more cumbersome
notation such as $n_{I(p,M)}$ would be more accurate where the
function $i = I(p,M)$ maps the $p^{th}$ populated bin to the
$i^{th}$ bin in the $M$-bin model.} This leads to an interesting
form for the marginal posterior probability for $M$
(\ref{eq:posterior-for-M}), since the function is no longer
dependent on the particular values of the data, just how many
instances of each discrete value was recorded, $n_1, n_2, \ldots,
n_P$.  Since these values do not vary for $M > M_{\Delta x}$, the
marginal posterior can be viewed solely as a function of $M$ with
a well-defined form
\begin{equation}
p(M | \underline{d}, I) \propto \biggl(\frac{M}{2}\biggr)^N
\frac{\Gamma\bigl(\frac{M}{2}\bigr)}{\Gamma\bigl(N+\frac{M}{2}\bigr)}
\cdot 2^N
\frac{\prod_{p=1}^{P}{\Gamma(n_p+\frac{1}{2})}}{\Gamma\bigl(\frac{1}{2}\bigr)^P},
\end{equation}
where the sum over $p$ is over populated bins only. Comparing this
to (\ref{eq:M>N}), the function on the right-hand side clearly
asymptotically approaches a value greater than one---so that its
logarithm increases asymptotically to a value greater than zero.

In cases where the value of this asymptote is greater than the
maximum value attained within the range $1 \leq M < M_{\Delta x}$,
the digitized structure of the data is a much more robust feature
than the statistical structure of the data itself before rounding
or truncation.  We explore some examples of this in the next
section.

\begin{figure}[t]
\centering
\includegraphics[width=0.85\textwidth]{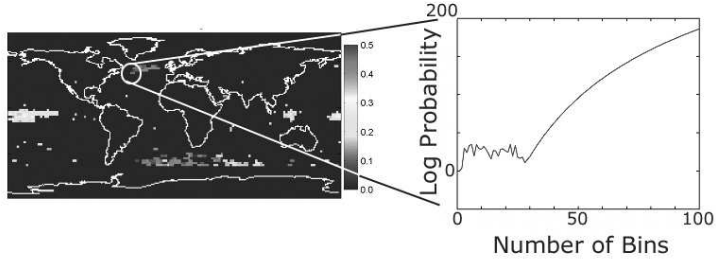}
\caption{During a mutual information study designed to examine the
effect of the El Ni\~{n}o Southern Oscillation (ENSO) on global
cloud cover (left) we found a region of pixels that caused
artifacts in our analysis, which relied on optimal histograms.
Careful examination revealed that the Percent Cloud Cover variable
in these regions was excessively rounded or truncated (right).
(Compare to Figure \ref{fig:1}D) In this case, it is likely that
there was more information present in the data than was originally
thought.}
\label{fig:3}       
\end{figure}

\subsection{Results}
To begin, let us refer to a previous example where 1000 data
points were sampled from a Gaussian distribution (Figures
\ref{fig:1}A and B).  In that example, the log probability
indicated that $M = 11$ would optimally describe the data set.  We
then took the same data, and rounded the values to the nearest
1/10th.  Modelling the density function using these excessively
rounded data values with a large number of bins shows a picket
fence effect (Figure \ref{fig:1}C) where the data are piled up on
their discrete values.  As predicted by the asymptotic analysis
above, the un-normalized log posterior probability increases
monotonically approaching an asymptote with a value greater than
zero (Figure \ref{fig:1}D).  Note that the behavior is very
different than that in the well-defined case shown in Figure
\ref{fig:2}.

In another study involving a mutual information analysis between
sea surface temperatures indicative of El Ni\~{n}o Southern
Oscillation (ENSO) and global cloud cover, we identified a small
region of pixels in the North Atlantic that seemed to be causing
artifacts in our analysis.  We were working with the Percent Cloud
Cover variable from the C2 data set from the International
Satellite Cloud Climatology Project (ISCCP)
\cite{Schiffer&Rossow:1983}, and found that for some areas, such
as the North Atlantic, the stored data values were excessively
rounded.  This effect can be easily seen in Figure \ref{fig:3}
where the log probability asymptotes as demonstrated in the
artificial case shown in Figures \ref{fig:1}C and D.  It is likely
that there is more information present in this variable than was
originally thought.

\begin{figure}[t]
\centering
\includegraphics[width=0.5\textwidth]{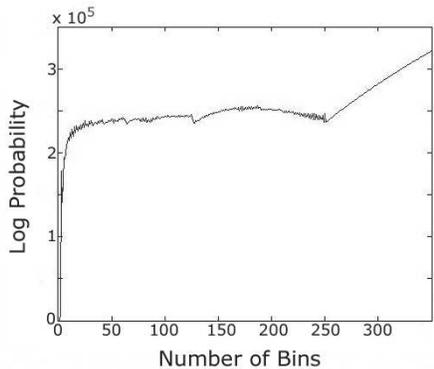}
\caption{The un-normalized log marginal posterior probability or
the number of bins for the surface reflectivity in a BRDF model
from a Level 2 MISR data product. Again this plot shows the
characteristic asymptotic behavior indicative of excessive
rounding or truncation.}
\label{fig:4}       
\end{figure}

The Multi-angle Imaging SpectroRadiometer (MISR) is an instrument
carried by the spacecraft Terra, which is part of NASA's Earth
Observing System.  Here we consider an example from a Level 2 MISR
data product, which describes the surface reflectivity in a
bidirectional reflectance factor (BRDF) model
\cite{Rahman+etal:1993}.  In this example, the data are stored as
8 bit unsigned integers (uint8), however since 253-255 are used
for overflow, underflow, and fill values, the stored data actually
range from zero to 252.  In Figure \ref{fig:4} we again show the
plot of the un-normalized log marginal posterior probability or
the number of bins, which after 252 bins shows the characteristic
asymptotic behavior indicative of excessive rounding or
truncation.  As in the previous case, information has been lost,
and unless it can be retrieved from a more primitive data product,
it cannot be regained.

\section{Conclusion}
We have demonstrated that a straightforward Bayesian method for
identifying the optimal number of bins in a piecewise-constant
density model demonstrates stereotypical behavior in the case
where the data have been excessively rounded or truncated.  By
``excessive'', we mean that the digitized structure of the data is
a much more robust feature than the statistical structure of the
original data. In such cases, an uninvertible transformation has
been applied to the data, and information has been irrevocably
lost.

We have demonstrated such excessive digitization in data from two
Earth observing satellite surveys.  In each case, it may be
desirable for researchers to know that information has been
discarded, even if to save transmission bandwidth or storage
space. However, it is not always clear that these decisions were
made wisely, nor is it clear that they should be made again in the
future.  For this reason, we expect that a simple tool developed
from the observations presented in this paper would find great use
in the scientific community both for engineers and scientists
working on the design aspects of a scientific instrument, and also
for researchers working on the data analysis.

\section{Acknowledgements}
This work was supported by the NASA Earth Science Technology
Office (ESTO) AIST-QRS-04-3010-T. The authors would also like to
thank William Rossow his assistance with the ISCCP data and the
mutual information research.

\index{paragraph}
%

%
%
%
%

%
%



\printindex
\end{document}